\documentclass[seceqn,secthm,TRL,twocolumn,10pt]{baltzer}
\usepackage{psfig}
\begin{document}
\begin{frontmatter}  
%
\title{Simulations of the kinetic friction due to adsorbed surface layers}
\author{Gang He and Mark O. Robbins
}
\address{Department of Physics \& Astronomy, Johns Hopkins University
\email{mr@pha.jhu.edu}}   
\runningauthor{He and Robbins}
\runningtitle{Kinetic friction from adsorbed layers}
\received{11 August 2000}
%
\begin{abstract}  
Simulations of the kinetic friction due to a layer of adsorbed
molecules between two crystalline surfaces are presented.
The adsorbed layer naturally produces friction that is consistent
with Amontons' laws and insensitive to parameters that are not
controlled in experiments.
The kinetic friction rises logarithmically with velocity as in many
experimental systems.
Variations with potential parameters and temperature follow variations
in the static friction.
This correlation is understood through analogy with the Tomlinson
model and the trends are explained with a hard-sphere picture.
\end{abstract}
\end{frontmatter}
\section{Introduction}

One of the major challenges for researchers in the field of nanotribology
is to uncover the molecular origins of macroscopic behavior.
For example, why do many systems obey Amontons' laws
that friction is proportional to normal load $L$ and
independent of the apparent geometric area of the surfaces $A_{\rm app}$?
Unfortunately many simple molecular scale models give results that are
qualitatively inconsistent with macroscopic measurements
\cite{tomlinson29,frenkel38,mcclelland89,hirano90,sorensen96,robbins96,aubry79,bak82,cieplak94,robbins00c,robbins00b}.
Indeed they suggest that static friction should almost
never be observed
\cite{sorensen96,robbins96,aubry79,bak82,cieplak94,robbins00c,robbins00b}!
A possible origin for such
discrepancies \cite{robbins96,he99,muser00,muser00a} is that the
models consider contacts between two bare surfaces.
Real surfaces almost always have an intervening layer of material
called "third bodies" \cite{godet84,berthier89,singer92a}.
This layer may be dust, grit, wear debris, or the
few angstroms of hydrocarbons and water that adsorb
rapidly on surfaces that are exposed to air.

In a recent paper He {\it et al.} \cite{he99} explored the effect of
adsorbed layers on the static friction $F_s$.
They showed that such layers naturally lead to a non-vanishing $F_s$
that exhibits much of the behavior seen in macroscopic experiments.
For example the calculated value of $F_s$ is insensitive to many
parameters that are not controlled in experiments,
including crystallographic alignment,
the thickness of the adsorbed layer, and the precise size
of the molecules.
More importantly, the simulation results provided a molecular basis for
Bowden and Tabor's phenomenological explanation of Amontons'
laws \cite{bowden86}.

Bowden and Tabor had
noted that $A_{\rm app}$ is
usually much larger than the area of intimate molecular
contact $A_{\rm real}$.
They pointed out that both Amontons' laws and many exceptions
to them could be explained {\em if} the local shear stress $\tau$
in the contacts increased linearly with the local pressure $P$:
\begin{equation}
\tau=\tau_0+\alpha P \ \ \ \ .
\label{eq:tau}
\end{equation}
Summing over $A_{\rm real}$ and dividing by load gives
a friction coefficient
\begin{equation}
\mu \equiv F/L = \alpha+\tau_0/ P .
\label{eq:mu}
\end{equation}
This is independent of load
if $P$ is constant or $\tau_0/P \ll \alpha$.
Later work showed that both elastic\cite{volmer97,greenwood66}
and plastic\cite{bowden86} models of surfaces with many contacts
yield a constant $P$ at high loads.
At low loads, $P$ decreases, explaining why many systems show an increase
in $\mu$ in this limit \cite{bowden86,rabinowicz65}.

He {\it et al.}'s work \cite{he99} showed that adsorbed layers lead to a
static friction that obeys Eq. \ref{eq:tau}.
However, most experiments measure the kinetic friction.
Although experimental values for the two quantities are often correlated,
it is not obvious that they should be.
Static friction is the threshold force that must be overcome
before one solid slides over another.
It arises because the two bodies have locked together into a
local energy minimum, and $F_{\rm s}$ is the force needed to
break them free from this minimum.
Kinetic friction $F_k(v)$ is the force needed to maintain sliding at
a given velocity $v$.
As such it is proportional to the work that must be done
due to energy dissipation during sliding.
Thus kinetic and static friction involve different phenomena
and they have very different values in many simple models
\cite{tomlinson29,frenkel38,mcclelland89,hirano90,sorensen96,robbins96,aubry79,bak82,cieplak94,robbins00c,robbins00b}.

In this paper we consider the effect of adsorbed layers on kinetic
friction.
We first show that the linear relation in
Eq. \ref{eq:tau} is obeyed for each velocity.
The intercept $\tau_0$ is relatively independent of velocity
below about 1 m/s.
The slope $\alpha$ shows a logarithmic dependence on velocity
over more than two decades in velocity.
This same functional dependence is observed in experiments, and
used in rate-state models of friction \cite{dieterich79,ruina83,gu84}.
We next examine the effect of interaction potentials, temperature,
and crystallographic alignment on kinetic friction.
The variations with these parameters are consistent with
a simple hard sphere model for lateral forces.
We conclude by considering the strong correlation between 
calculated values of $F_k$ and $F_s$.
A qualitative explanation for this correlation is suggested based on
results for the Tomlinson model \cite{tomlinson29},
and this analogy is confirmed
by detailed analysis of the dynamics of individual molecules.

The following section describes the model we use for the surfaces
and the intervening molecular layer.
The next section presents results, and the final section provides
a summary and conclusions.

\section{Computational Methodology}

Since our goal is to establish general mechanisms, we use
relatively simple interaction potentials in our simulations.
These allow us to quickly span a wide range of sliding velocities,
system sizes, geometries and interactions.
They also allow treatment of longer time and length scales than
more detailed potentials.

We find similar behavior for both crystalline and disordered walls
\cite{he99,muser00,muser00a}.
For the results described below, each solid contains discrete atoms
forming the (111) surface of an fcc crystal.
The coordinate system is chosen so that the surfaces lie in the $x-y$ plane,
and periodic boundary conditions are applied in these directions
to minimize edge effects.
Note that the periodic boundary conditions also prevent us from considering
perfectly incommensurate systems.
However, we find that surfaces exhibit incommensurate behavior as soon
as the common period becomes longer than a few lattice
constants \cite{muser00}.

The two opposing surfaces have approximately the same nearest-neighbor
spacing $d_{nn}$, and
the effect of commensurability is explored by rotating the top surface
by an angle $\theta$ about the $z$-axis.
A discrete set of angles is studied that allow both surfaces to retain
their hexagonal symmetry.
Except where noted, the bottom wall has $d_{nn}=1.2\sigma$.
The value of $d_{nn}$ for the top wall is adjusted slightly
($\le 2.1$\%) in order to satisfy the periodic boundary conditions.
No correlation between the size of this adjustment and the
calculated friction was seen at the level of the statistical fluctuations.

Atoms in the bottom solid are fixed in place, and
atoms in the top solid translate together as a rigid body of mass $M$.
Previous work shows that including elastic deformations of the solids
produces a slightly lower static friction \cite{he99,muser00,he00},
and we find a similar reduction in the kinetic friction.
A constant downward pressure is applied to the top solid, and its height
varies in response to this external pressure and molecular interactions.
A constant lateral force or velocity is applied along the nominal sliding
direction, usually $\hat{x}$.
The wall is free to move in the other lateral direction,
i.e. the external force is zero in this direction.
The kinetic friction is determined by imposing a constant lateral
velocity $v$, and calculating the average opposing force.
This average is performed over a displacement of at least 25$d_{nn}$
in order to reduce statistical fluctuations.
The static friction is obtained by applying a constant lateral force,
and determining the value needed to initiate sliding \cite{he99}.

The layer of molecules between the solids is described with
a bead-spring model.
Previous studies have shown that this model yields realistic dynamics for
polymer melts \cite{kremer90},
and shown how to map between it and detailed chemical models of
polymers \cite{tschop98a,tschop98b}.
These detailed models take at least
an order of magnitude more computer time.

Each molecule contains $n$ spherical monomers of mass $m$.
We considered simple spherical molecules $n=1$, and
short chains that model airborn hydrocarbons, $n=3$ and 6.
All monomers interact through a truncated Lennard-Jones potential
\cite{allen87}:
\begin{equation}
V_{LJ}(r)=4\epsilon \left[(\sigma/r)^{12}-(\sigma/r)^6\right] \ \ \ {\rm for}
\ r < r^c,
\end{equation}
where $r$ is the distance between molecules and the potential is
zero for $r>r^c$.
Adjacent monomers along a chain are coupled by an additional potential
that prevents chains from crossing or breaking:
\begin{equation}
V_{CH}(r) = -(1/2) k {R_o^2} \ln [1-(r/R_o)^2]\ \ \ \ \ ,
\end{equation}
where $R_o = 1.5\sigma$ and $k = 30\epsilon/\sigma^2$ \cite{kremer90}.
The parameters $\epsilon$, $\sigma$, and
$t_{LJ}\equiv(\sigma^2/m\epsilon)^{1/2}$
are characteristic energy, length, and time scales, respectively.
These characteristic scales are used to normalize other quantities.
Values that would be representative of hydrocarbons are
$\epsilon \sim 30$meV, $\sigma \sim 0.5$nm and
$t_{LJ} \sim 3$ps \cite{kremer90}.
The unit of pressure, $\epsilon \sigma^{-3}$, corresponds to about 40MPa.

Wall atoms interact with monomers through a Lennard-Jones potential
with a different cutoff, $r^c_{wf}$,
and different energy and length scales, $\epsilon_{wf}$ and $\sigma_{wf}$,
respectively.
This allows us to increase or decrease the amount of adhesion,
and the effective surface corrugation of the walls.
In most cases we considered purely repulsive interactions:
$r^c=2^{1/6}\sigma$ and $r^c_{wf}=2^{1/6}\sigma_{wf}$.
In this limit, atoms on opposing walls are far enough apart
that they do not interact.
As shown below,
the main effect of increasing the interaction range is to add
an extra adhesive pressure between the walls.

All of our runs are done in a constant temperature ensemble.
We have performed extensive tests on the effects of
thermostats \cite{robbins00c,thompson90a,stevens93,smith96}.
In the appropriate limits, thermostats prevent a gradual rise in
temperature due to frictional dissipation without having a direct
influence on the friction force.
To ensure that one is in these limits, the thermostat must be
applied normal to the sliding direction and be
sufficiently weak,
and the velocity must be low ($\leq 0.1\sigma/t_{LJ}$).
At high velocities, the structure and other properties of the film
may be affected \cite{manias94,khare96}.
In the following simulations,
a Langevin thermostat with damping rate $0.4 t_{LJ}^{-1}$
is coupled to the equations of motion in both directions
normal to the sliding direction \cite{thompson90a,grest86}.

The equations of motion are integrated using a fifth-order Gear
predictor-corrector algorithm with a time step $dt=0.005\sigma/t_{LJ}$.
Unless otherwise noted, the area of a periodic unit cell is
${\cal A}=720 \sigma^2$ and the mass of the wall is
$M/m =0.8 {\cal A}/\sigma^2$.
Previous studies show that finite-size effects are very small at
this system size \cite{he00}.
Most simulations are done at temperature $T=0.7k_B/\epsilon$,
where $k_B$ is Boltzmann's constant.

\section{Results}
\subsection{Variation with pressure and velocity}

\begin{figure}
\begin{picture}(000,205)
\put(0,0){\psfig{figure=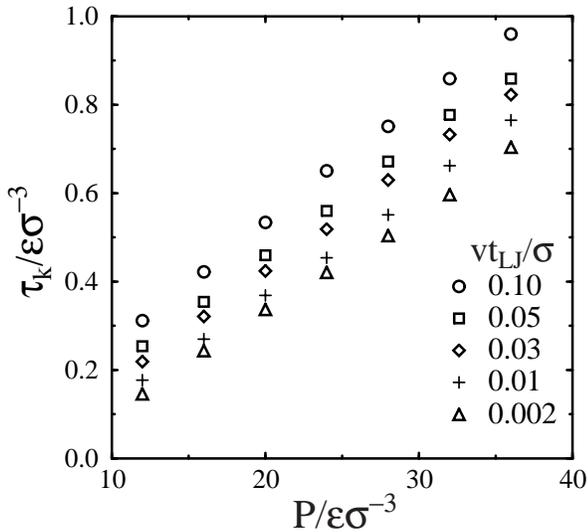,width=80mm}}
\end{picture}
\caption {Kinetic shear stress $\tau_k$ as a function of pressure
for the indicated velocities.
Velocities are in units of $\sigma/t_{LJ} \sim 100$m/s,
and the unit of stress and pressure is $\epsilon \sigma^{-3} \sim 40$MPa.
For these runs $\theta=30^\circ$, $\sigma_{wf}=\sigma$,
$\epsilon_{wf}=\epsilon$, $n=6$, $d_{nn}=1.2\sigma$, and the number
of monomers is equal to the number of atoms on the surface of the
bottom wall.
The data for each velocity lies on a straight line.
We confirmed that linearity extended
up to $P/\epsilon\sigma^{-3}=100$ for the highest velocities.
}
\label{fig:tau_p_v}
\end {figure}

Figure \ref{fig:tau_p_v} shows the pressure dependence of the kinetic
friction at the indicated velocities for an incommensurate system,
$\theta=30^\circ$.
At each velocity the shear stress $\tau_k$ rises linearly with pressure
over the entire range of pressures.
Using realistic values of $\epsilon$ and $\sigma$, the highest pressures
correspond to about 1.5GPa.
The linear relation continues up to $P=100\epsilon \sigma^{-3}$
in the cases we have tested.
However, the larger forces at these pressures require a reduction in
the integration time step, making the calculations much slower.

\begin{figure}[h]
\begin{picture}(000,200)
\put(0,0){\psfig{figure=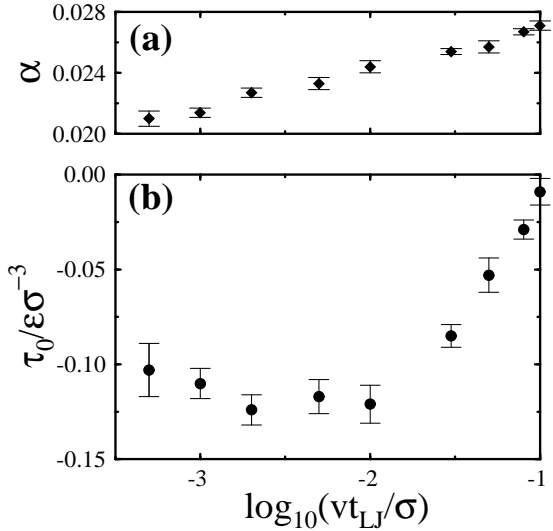,width=75mm}}
\end{picture}
\caption {Plot of (a) $\alpha$ and (b) $\tau_0$ against the logarithm
of the velocity for the system of Fig. \ref{fig:tau_p_v}.
}
\label{fig:kv90}
\end{figure}

Values of $\alpha$ and $\tau_0$ were obtained from linear fits to
data like that shown in Fig. \ref{fig:tau_p_v}.
In each case the correlation coefficient was greater than .999.
The results are plotted against the logarithm of velocity in
Figure \ref{fig:kv90}.
The value of $\alpha$ increases as the logarithm of velocity
over the entire range of Fig. \ref{fig:kv90}
-- a factor of 200 in $v$.
In contrast, the
value of $\tau_0$ becomes independent of velocity for
$v < 0.01 \sigma/t_{LJ} \sim 1$m/s.
It is negative because these simulations used a purely
repulsive potential $r^c/\sigma = 2^{1/6}$.
We show later that including adhesive interactions leads to positive values
of $\tau_0$.

A logarithmic variation of friction with velocity is also seen
in many experimental systems
\cite{dieterich79,ruina83,gu84}.
Measured friction coefficients fit a ``rate-state'' equation:
\begin{equation}
\mu = \alpha + \tau_0/P = \mu_0 + A \ln (v/v^0) + B \ln (\Theta/\Theta^0) ,
\label{eq:state}
\end{equation}
where $\Theta$ is a variable that describes the
state of the system, and
$\mu_0$ is the friction at the reference velocity $v^0$ and state $\Theta^0$.
In most cases $\Theta$ is treated as a phenomenological parameter
whose evolution is described by:
\begin{equation}
d\Theta/dt = 1-\Theta v/D_c
\label{eq:Theta}
\end{equation}
where $D_c$ has units of length.
Dieterich and Kilgore have shown that
changes in $\Theta$ are proportional changes in the true area of
contact between rough surfaces
and that $D_c$ corresponds to the typical diameter of contacts
\cite{dieterich96}.

Our simulations give the dependence of friction on $v$ at fixed
contact area \cite{footx}
and thus the slope of Fig. \ref{fig:kv90}(a) gives
$A=0.00110\pm0.00005$.
This is smaller than typical experimental values for glass or rocks
where $A$ is .005 to .015.
However, one might expect $A$ to increase with $\alpha$.
While our values of $\alpha$ are comparable to values for flat
crystalline surfaces like mica\cite{granick94,gee90}
or MoS$_2$\cite{singer92a}, they are much smaller than values for rocks.
The ratio $A/\alpha$ is about $.05$ for our system, and varies from
.008 to .025 for rocks.
We do not know of any measurements of $A$ for mica or MoS$_2$, but
it would be interesting to see how results for these systems
compare to our simulations.

\subsection{Variation of $\mu$ with interaction potential, $T$ and geometry}

\begin{figure}
\begin{picture}(000,200)
\put(0,0){\psfig{figure=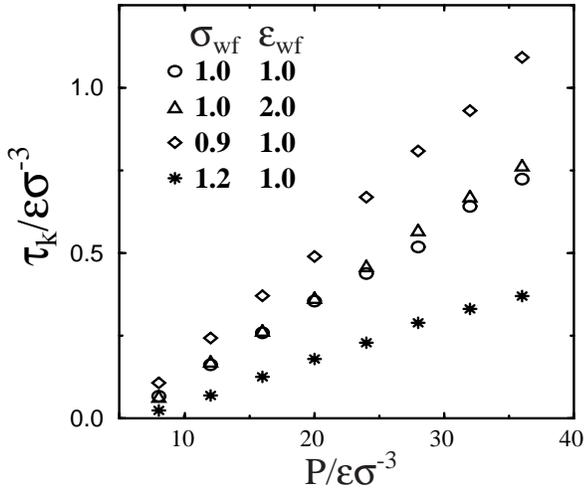,width=80mm}}
\end{picture}
\caption {Plot of shear stress against pressure at $v=0.005\sigma/t_{LJ}$
for the indicated values of $\sigma_{wf}$ and $\epsilon_{wf}$.
Other parameters are the same as for Fig. \ref{fig:tau_p_v}.
}
\label{fig:pot}
\end{figure}

A variety of interaction potentials and geometries have
been considered to determine which factors
influence $\alpha$ and $\tau_0$ and which leave them unchanged.
As shown in Fig. \ref{fig:pot}, doubling the interaction between
wall atoms and fluid monomers has almost no effect on the friction.
In contrast, the friction increases rapidly with the ratio of
$d_{nn}$ to $\sigma_{wf}$.
For $d_{nn}=1.2\sigma$ we find $\alpha$ rises from 0.012
to 0.034 as $\sigma_{wf}$ drops from 1.2 to 0.9.
The value of $\tau_0$ increases in magnitude from -0.07 to -0.16.
Although not shown, increasing $d_{nn}$ at fixed $\sigma_{wf}$
also increases $\alpha$.

\begin{figure}
\begin{picture}(000,85)
\put(0,0){\psfig{figure=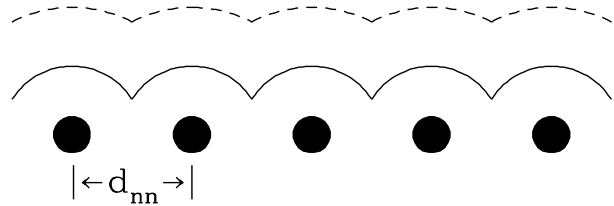,width=80mm}}
\end{picture}
\caption {Sketch of surface of closest approach for
$D/d_{nn}=0.7$ (solid line) and 1.3 (dashed line).
The lateral force required to lift monomers up the ramps
that these surfaces represent is given by the slope times
the normal force.
}
\label{fig:sketch}
\end{figure}

These trends in $\alpha$ can be understood from a simple geometrical
picture.
At the pressures of interest, the interactions are dominated by
the repulsive $r^{-12}$ term in the Lennard-Jones potential.
Monomers and wall atoms can be thought of as hard spheres
with an effective diameter $D$ that is determined by the separation
$r$ at which the Lennard-Jones force balances the force from the applied
pressure.  This picture gives a diameter
$D \propto \sigma_{wf} (\epsilon_{wf}/\sigma_{wf} P)^{-1/13}$,
that varies rapidly with $\sigma_{wf}$ but very little with $\epsilon_{wf}$
or $P$.
As illustrated in Fig. \ref{fig:sketch},
monomers can penetrate deep between surface atoms
when $D/d_{nn}$ is small.
This increases the friction because monomers must be lifted up
the ramp formed by the surface of closest approach in order to translate.
The required lateral force is the normal load times
the slope of the surface which increases as $D/d_{nn}$ decreases.
Thus the hard-sphere model explains the linear dependence of $\tau$
on pressure as well as the insensitivity to $\epsilon_{wf}$
and the variation with $\sigma_{wf}$.

The simulations just described were for strictly repulsive potentials
$(r^c_{wf}=2^{1/6}\sigma_{wf})$,
and one may wonder whether the hard-sphere picture is more generally
applicable.
We have done simulations with $r^c_{wf}=2.2\sigma_{wf}$ to test this.
When direct wall/wall interactions are neglected, we find that
the kinetic friction is fit by $\alpha = 0.0211\pm 0.0003$ and
$\tau_0=+0.007\pm .008$
over a pressure range from 0 to 36$\epsilon \sigma ^ {-3}$ for
the parameters of Fig. \ref{fig:tau_p_v}.
This value of $\alpha$ is only 8\% smaller than for a purely repulsive
potential.
In contrast, $\tau_0$ increases substantially because the attraction
between monomers and wall atoms increases the effective pressure
holding them together.
Adding direct wall/wall interactions leads to a slightly larger adhesive
pressure, and a slightly larger value of $\tau_0$.

\begin{figure}
\begin{picture}(000,240)
\put(0,0){\psfig{figure=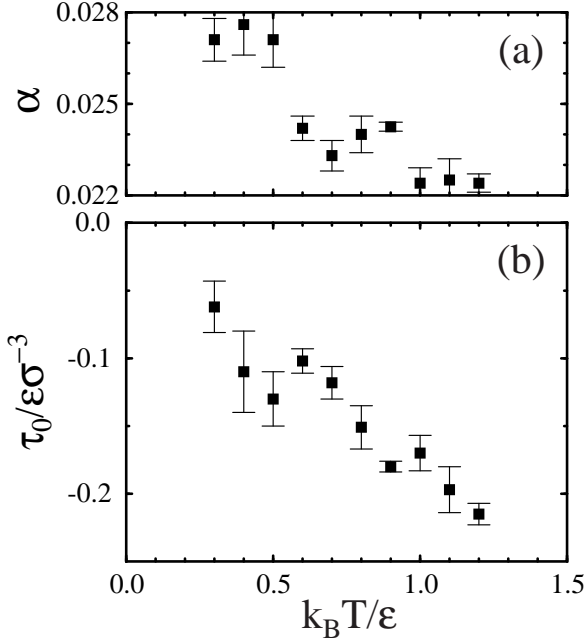,width=80mm}}
\end{picture}
\caption {Variation of (a) $\alpha$ and (b) $\tau_0$ with
temperature at $v=0.005\sigma/t_{LJ}$
for the system of Fig. \ref{fig:tau_p_v}.
}
\label{fig:temp}
\end{figure}

The effect of temperature on $\alpha$ is also fairly weak.
So far we have shown results for $T=0.7 \epsilon/k_B$, which
is near the triple point for spherical Lennard-Jones molecules,
and roughly 30\% above the glass transition temperature for chain
molecules.
This temperature was chosen because many airborn hydrocarbons are
above their melting point at room temperature.
Figure \ref{fig:temp} shows the variation of $\alpha$ and $\tau_0$
with temperature at $v=0.005\sigma/t_{LJ}$
for the system of Figs. \ref{fig:tau_p_v}
and \ref{fig:kv90}.
The value of $\alpha$ changes only weakly with temperature,
falling about 20\% from $k_B T/\epsilon=0.3$ to 1.2.
The high end of this temperature range is more than twice the
glass transition temperature for chain molecules and close
to the critical temperature for spherical molecules.

The weak variation of $\alpha$ with $T$ is consistent with our geometrical
picture, since the hard-sphere diameter is insensitive to temperature.
In contrast,
the value of $\tau_0$ is roughly proportional to temperature
over the studied range.
Equation \ref{eq:tau} implies that the friction vanishes
for $P < - \tau_0/\alpha$.
Of course this can not happen, but there is a qualitative
change in the type of friction as the pressure drops below
$-\tau_0/\alpha$.
At high pressures we observe static friction and a slightly
smaller kinetic friction that rises logarithmically with velocity.
For $P < -\tau_0/\alpha$ the static friction vanishes, and the
kinetic friction rises from zero as a power of the velocity.

This change in frictional behavior reflects a change in the
state of the film that has been studied previously in
multi-layer systems
\cite{bitsanis90b,thompson92,thompson95,landman96,manias96,robbins00a}.
At high pressures the molecules lock into a glassy state,
while at low pressures they diffuse freely and act like
a viscous lubricant.
Fig. \ref{fig:temp} implies that the pressure required to lock
molecules into a glassy state rises linearly with $T$.
This result is specific to our choice of purely repulsive
potentials ($r^c=2^{1/6}\sigma$).
Adding an attractive tail to the potential shifts the entire
temperature dependent curve upwards, and gives glassy behavior
at lower (even negative) pressures.
Note that earlier results on multi-layer films
indicated that $\tau$ was independent of velocity in the glassy
state \cite{thompson95}.
However,  these results covered a relatively small range of $v$
and are not inconsistent with the logarithmic variation seen in
Fig. \ref{fig:kv90}.

\subsection{Relation to static friction}

The changes in kinetic friction with temperature, interaction potential,
geometry and other factors are very similar to what we have observed
for static friction \cite{he99,he00}.
Indeed the ratio between the value of $\alpha_k(v)$
at a given velocity
and the value of $\alpha_s$ for static friction is relatively constant.
For example, over the range of temperatures and potential
parameters shown in Figures \ref{fig:pot} and \ref{fig:temp},
$\alpha_k( v=0.005 \sigma/t_{LJ})$
is between 75 and 85\% of $\alpha_s$.
This variation is comparable to our statistical
uncertainties.

\begin{figure}
\begin{picture}(000,205)
\put(0,0){\psfig{figure=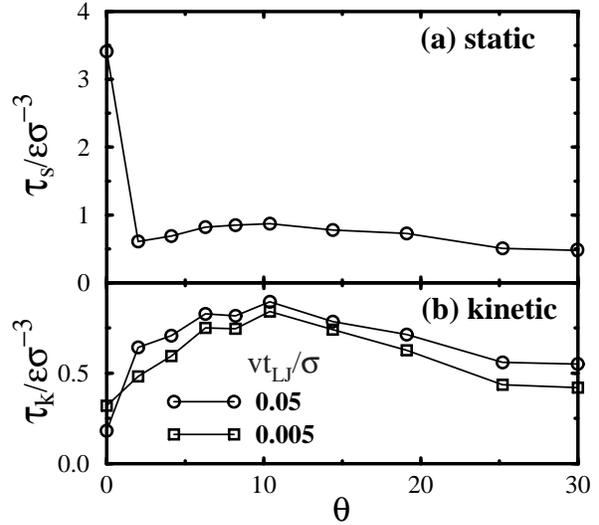,width=80mm}}
\end{picture}
\caption {Variation of (a) static and (b) kinetic
friction with $\theta$ at $P=24\epsilon \sigma^{-3}$.
Error bars in both panels are about 5\%.
For these simulations the area of the unit cell is $2880\sigma^2$,
and the number of monomers is equal to the number of atoms on the
surface of the bottom wall.
}
\label{fig:sko}
\end{figure}

Figure \ref{fig:sko} illustrates the correlation between static
and kinetic friction as $\theta$ is varied.
Only the range $\theta =0$ to 30$^\circ$ is shown, since the values
at other $\theta$ are related by symmetry.
The simulation data in previous figures was for $\theta=30^\circ$
and the commensurate case corresponds to $\theta=0^\circ$.
We confirmed that the friction is linear in $P$ for several $\theta$,
and show results for a single, relatively high pressure
($P=24 \epsilon \sigma^{-3}$) to minimize
the number of calculations, and emphasize the contribution from $\alpha$.
Data for a monolayer film is shown because it gives the largest variation in
friction with $\theta$ and thus the best test of correlations
between $F_k$ and $F_s$ \cite{footx2}.
As $\theta$ changes from 2$^\circ$ to 30$^\circ$, $F_k$ varies by
almost a factor of two, yet the ratio $F_k/F_s$ stays constant
within our statistical uncertainties.
For $v=0.005\sigma/t_{LJ}$ we find $F_k/F_s = 0.88\pm 0.10$,
while $F_k/F_s = 1.05 \pm 0.10$ for $v=0.05\sigma/t_{LJ}$.

The correlation between $F_k$ and $F_s$ is lost when the walls
become commensurate.
As $\theta$ decreases from 2$^\circ$ to $0^\circ$ there is a sharp
increase in $F_s$ by about a factor of seven,
and a drop in $F_k$ by about a factor of two.
In addition, there is a qualitative change in the velocity dependence
of the kinetic friction:
$F_k$ decreases with velocity in the commensurate case, but
increases with $v$ for incommensurate walls.

The observed correlation between static and kinetic friction
is somewhat surprising given that they reflect different physical
effects.
As noted in the introduction, the static friction is related to
the force needed to pull the system out of a local potential
energy minimum, while the kinetic friction is related to the
energy dissipated during sliding.
Nevertheless,
the above comparisons of $F_k$ and $F_s$ indicate that the two arise
from similar processes when the walls are incommensurate.

\subsection{Comparison to the Tomlinson model}

Some insight into the relation between static and kinetic
friction is provided by a simple one-dimensional ball and spring model
called the Tomlinson model
\cite{tomlinson29,mcclelland89,robbins00c,robbins00b}.
This model considers two bare surfaces in contact.
The bottom surface is modeled by a sinusoidal potential with
amplitude $V_0$ and period $d_{nn}$.
Atoms in the top surface are treated as independent, 
overdamped oscillators that
are coupled to the center of mass $X_{CM}$ by a spring of stiffness $k$.

The behavior of the Tomlinson model depends on the dimensionless parameter
$\lambda \equiv 4\pi^2V_0/kd^2_{nn}$ that measures the relative
stiffness of the potential and spring.
When $\lambda$ is small, the spring is so stiff that each atom
has only one metastable state at any $X_{CM}$.
In this limit atoms move smoothly over the periodic
potential from the bottom wall.
This nearly reversible motion leads to little dissipation,
and $F_k$ goes to zero linearly with the center of mass
velocity \cite{mcclelland89,robbins00c,robbins00b}.
When $\lambda$ is large, each atom can get locked in many metastable
states at any center of mass position.
Each atom starts in one such state.
As $X_{CM}$ increases, that state eventually becomes unstable and
the atom pops rapidly to the next state.
During these pops the system is far from equilibrium and a
substantial amount of energy is dissipated.
The sequence of pops and the amount of energy dissipated during
a given wall displacement $\Delta X_{CM}$ is independent of
wall velocity in the limit $V_{CM}\rightarrow 0$.
Since the dissipated energy equals $F_k(v) \Delta X_{CM}$,
the kinetic friction approaches a constant
value, $F_k (0)$, in this limit.
One can show that $F_k(0)$ approaches the maximum
force exerted by the periodic potential $F_{max}\equiv 2\pi V_0/d_{nn}$
times the number of atoms $N$ as $\lambda \rightarrow \infty$
\cite{mcclelland89,robbins00c,robbins00b}.

The static friction in the Tomlinson model depends on whether the
walls are commensurate or
incommensurate \cite{robbins00c,robbins00b}.
In the commensurate case all atoms feel the same periodic
potential at each $X_{CM}$.
The maximum force that must be overcome as they advance by a period is
$F_s=NF_{max}$.
The ratio $F_k/F_s$ vanishes for $\lambda < 1$ and approaches
unity as $\lambda \rightarrow \infty$.
In the incommensurate case $F_s=0$ for $\lambda < 1$.
For $\lambda > 1$, the static friction is equal to the low
velocity kinetic friction.
This link between $F_s$ and $F_k$
can be seen by calculating $F_k(0)$ from instantaneous forces
rather than the dissipated energy.
At very low velocities almost all of the atoms are stuck in a
metastable minimum at any instant in time.
The force resisting motion, $F_k(0)$, comes from the force exerted
by the periodic potential on these atoms.
One can show that summing this force or calculating
the energy dissipated during atomic pops gives the same answer
for $F_k$ \cite{robbins00c,robbins00b}.
For incommensurate walls the distribution of metastable states
and thus the lateral force is independent of position.
Hence
the static friction equals the force $F_k(0)$ needed to advance
the wall at a very low velocity.

In most cases two bare surfaces will be incommensurate and
the interactions between them will be weaker than their internal
interactions.
This corresponds to the case $\lambda <1$ and one expects the
static friction and low velocity kinetic friction to vanish
\cite{mcclelland89,hirano90,sorensen96,robbins96,aubry79,bak82,cieplak94,robbins00c,robbins00b}.
In contrast, the layer of adsorbed molecules included in our calculations
acts like a very compliant solid with $\lambda >1$.
As expected from the Tomlinson model, this produces a finite static
friction and a comparable kinetic friction.
The lateral force is predominantly determined by the hard sphere
interactions which can be expressed as the normal load times
the local slope of the surface of closest approach
(Figure \ref{fig:sketch}).
Since the surface is nearly independent of load and other parameters,
the lateral force is proportional to load and Amontons' laws
are obeyed.

We have confirmed the connection between our results and the Tomlinson
model by analyzing the motion of individual monomers
at low velocities.
The number of particles that move by more than a distance $c$ over
a time interval of $2 t_{LJ}$ is calculated as a function of $c$.
This time interval is chosen because it is longer than the velocity-velocity
autocorrelation time and the typical period of oscillations of
individual monomers about their metastable minima.
We focus on the case where the film is in a glassy state and
static friction is observed ($P > -\tau_0/\alpha$).

In equilibrium ($v=0$), the distribution of
monomer displacements is consistent with thermal
motion about metastable minima plus extremely
slow diffusion of monomers between metastable minima.
When the top wall is advanced at a low velocity,
almost all of the monomers are locked in local metastable minima
during any short time interval.
As in the Tomlinson model,
they provide almost all of the force on the top wall and thus
determine $F_k$.
Monomers that advance by significantly more than thermal displacements
are far fewer in number and contribute a smaller force per atom to $F_k$.
For example, with $v=0.0005\sigma/t_{LJ}$ at $k_B T/\epsilon=0$
only 0.3\% of the monomers move by more than 0.12$\sigma$ in 2$t_{JL}$.
These monomers contribute less than 0.1\% to the lateral force on the walls,
and the percentage decreases as the velocity is lowered.

When the walls are commensurate, the distribution of metastable
states depends on the relative positions of the walls.
The number of atoms that pop between metastable states is
a periodic function of the wall displacement.
No pops occur when the walls are aligned directly above each other,
and the total energy of the system is minimized in this configuration.
The number of pops rises rapidly as the walls are moved out of
alignment, and there are pronounced clusters of pops at unfavorable
wall positions.
The static friction is the maximum force exerted on the top wall
at any position,
while the kinetic friction represents an average over all wall positions.
As a result, the two can be very different and the ratio depends
on many factors including the length of the chains, number of chains, etc..

When the walls are incommensurate, the distribution of metastable
states is independent of wall position.
Thus the kinetic and static friction are correlated.
The analogy to the Tomlinson model would suggest that the kinetic
and static friction should be equal in this limit.
As illustrated above, this ratio depends on the velocity at which
$F_k$ is calculated and it may also depend on how $F_s$ is calculated.
Quoted results for $F_s$ were obtained by determining the minimum force
needed to produce a displacement of $\sigma$ in a time $100 t_{LJ}$.
One might thus expect the results to correspond to an effective
velocity of order 0.01$\sigma/t_{LJ}$, and the static friction is
indeed between the values of $F_k$ for $v=0.005$ and 0.05$\sigma/t_{LJ}$.
Rate-state models
(Eqs. \ref{eq:state} and \ref{eq:Theta}) also imply that the
static friction depends on the time spent at zero velocity before
a lateral force is applied.
We are currently exploring the effect of waiting time on static
friction and its relation to the velocity dependent kinetic friction.

\section{Summary and Conclusions}

We have examined the kinetic friction of surfaces separated by
a layer of adsorbed molecules as a function of
velocity, temperature, interaction potentials and surface alignment.
The results are qualitatively consistent with many macroscopic measurements,
and can be understood by analogy to simple model systems.

One of our main results is that the kinetic friction rises
logarithmically with velocity.
This dependence is commonly observed in experimental systems and
is often associated with thermal activation
\cite{dieterich79,ruina83,gu84}.
Our analysis of the dynamics of individual monomers is consistent
with this idea.
We found that the kinetic friction was given by the force exerted
by monomers that were trapped in metastable states.
A simple argument shows that this force will decrease as
$v$ decreases.

The force from the top wall on a monomer in a given local potential
well is zero at the bottom of the well.
As it is displaced away from the minimum the force rises.
When the force reaches its maximum value the monomer becomes unstable
and pops to a new well.
Since atoms that exert the largest force are closest to
becoming unstable, they are the ones that are most likely
to be thermally activated.
When they hop to a new well, they contribute a smaller lateral force.
As $v$ decreases the time for thermal activation increases as $\sigma/v$.
Thus more and more of the large contributions
to $F_k$ are lost due to thermal activation.
As a result, $F_k$ drops as $v$ decreases.

Another important result is that adsorbed layers naturally
lead to a kinetic friction that satisfies Eq. \ref{eq:tau}.
He {\it et al.} had previously established this relation for
the static friction \cite{he99}.
Thus adsorbed layers provide a molecular basis for Bowden and Tabor's
explanation of Amontons' laws in both contexts.

Changes in the kinetic friction with interaction potential
and temperature closely parallel changes in the static friction.
Both $F_k$ and $F_s$ are insensitive to parameters that are not
controlled in typical experiments.
Examples include wall orientation ($\theta$), the number and precise
size of adsorbed molecules, and even the strength of adsorption
($\epsilon_{wf}$).
This insensitivity
is consistent with the fact that variations in the measured
friction between nominally dry surfaces are relatively small ($\sim 25$\%).
In contrast, simulations and analytic studies of bare surfaces
show dramatic variations with surface orientation
\cite{hirano90,sorensen96,robbins96,aubry79,bak82,cieplak94,robbins00c,robbins00b}.

The linear relation between shear stress and pressure in Eq. \ref{eq:tau}
can be understood from a simple hard sphere picture (Fig. \ref{fig:sketch}).
The wall atoms create a surface of closest approach that adsorbed
molecules must climb up in order to move laterally.
The lateral force and normal force are related by the local slope
of this surface, and $\alpha$ represents some average of the slope.
Since $\epsilon_{wf}$ has little affect on the hard sphere diameter,
changing it has little affect on $\alpha$.
In contrast, changing the ratio $\sigma_{wf}/d_{nn}$ has a direct
effect on both the slope and $\alpha$
(Figs. \ref{fig:pot} and \ref{fig:sketch}).

The simple hard sphere model also
explains the insensitivity of $\alpha$ to the number of adsorbed
molecules per unit surface area $\rho$ \cite{he99,he00}.
Increasing the density of molecules spreads the normal pressure $P$
and lateral stress $\tau$ over more molecules.
However, the ratio between them, $\alpha$, is still determined by the slope
of the surface of closest approach.
This surface is determined solely by the solids when the film is
less than a monolayer thick.
One only expects variations in $\alpha$ when the film becomes
thicker than a monolayer and can shear internally.
Preliminary studies of the thickness dependence do show such effects
as the thickness increases to two or three layers.

The close connection between our results for kinetic and static friction
is typical of experiments on nominally dry surfaces
and surfaces that are purposely coated with thin hydrocarbon films
\cite{persson98}.
We discussed how
this connection can be understood from results for the simple
Tomlinson model.
Although the ratio $F_k(v)/F_s$ is relatively constant for a
given velocity, there are noticeable changes with $v$ due
to the logarithmic variation in $F_k(v)$.
We expect that the value of $F_s$ will also vary logarithmically with
the measurement time, and this remains an avenue for future research.

Another important focus of future research will be to improve
the quantitative comparison between theory and experiment.
The simulations described here use
simple interaction potentials and surface
geometries that capture qualitative features of the friction
due to adsorbed layers.
More detailed models of molecular structure and bonding,
as well as surface roughness and elasticity will be needed to
allow direct comparison to experiment.

\acknowledgements
The authors wish to thank
Drs. M. H. M\"user and P. M. McGuiggan for useful discussions.
Support from National Science Foundation Grant No. DMR-0083286
and from Intel Corporation through the donation of workstations
is gratefully acknowledged.


\begin{thebibliography}{00}
\bibitem{footx}
Note that while increasing velocity always increases the friction at fixed
  $\Theta$, the steady state kinetic friction may increase or decrease. From
  Eq. \ref{eq:Theta} the steady state of the system corresponds to
  $\Theta=D_c/v$, giving $\mu=\mu_0+(A-B)\ln(v/v^0)$. This decreases with
  increasing velocity if $A < B$.
\bibitem{footx2}
For submonolayer films the variation in $\alpha$ with $\theta$ is less than
  $\pm 15\%$.
\end{thebibliography}

\end{document}